
 \def\LOADED{\relax}

 \ifx\BoxedEPSFLoaded\LOADED
  \immediate\write16{}
  \immediate\write16{  BoxedEPSF macros already defined!}
  \immediate\write16{}
  \endinput
 \fi

 \let\BoxedEPSFLoaded\LOADED

 \chardef\CatAt\the\catcode`\@
 \chardef\CatColon\the\catcode`\:
 \catcode`\@=11
 \catcode`\:=12

 \let\wlog@ld\wlog
 \def\wlog#1{\relax}

 \newif\ifIN@
 \newdimen\XShift@ \newdimen\YShift@
 \newtoks\Realtoks

  %
 \newdimen\Wd@ \newdimen\Ht@
 \newdimen\Wd@@ \newdimen\Ht@@
 \newdimen\TT@
 \newdimen\LT@
 \newdimen\BT@
 \newdimen\RT@
 \newdimen\XSlide@ \newdimen\YSlide@
 \newdimen\TheScale  
 \newdimen\FigScale  
 \newdimen\ForcedDim@@

 \newtoks\EPSFDirectorytoks@
 \newtoks\EPSFNametoks@
 \newtoks\BdBoxtoks@

 \newif\ifNotIn@
 \newif\ifForcedDim@
 \newif\ifForcedHeight@
 \newif\ifPSOrigin

 \newread\EPSFile@

 \newif\ifIN@\def\IN@{\expandafter\INN@\expandafter}
  \long\def\INN@0#1@#2@{\long\def\NI@##1#1##2##3\ENDNI@
    {\ifx\m@rker##2\IN@false\else\IN@true\fi}%
     \expandafter\NI@#2@@#1\m@rker\ENDNI@}
  \def\m@rker{\m@@rker}

  \newtoks\Initialtoks@  \newtoks\Terminaltoks@
  \def\SPLIT@{\expandafter\SPLITT@\expandafter}
  \def\SPLITT@0#1@#2@{\def\TTILPS@##1#1##2@{%
     \Initialtoks@{##1}\Terminaltoks@{##2}}\expandafter\TTILPS@#2@}


  \newtoks\Trimtoks@

 \def\ForeTrim@{\expandafter\ForeTrim@@\expandafter}
 \def\ForePrim@0 #1@{\Trimtoks@{#1}}
 \def\ForeTrim@@0#1@{\IN@0\m@rker. @\m@rker.#1@%
     \ifIN@\ForePrim@0#1@%
     \else\Trimtoks@\expandafter{#1}\fi}

  \def\Trim@0#1@{%
      \ForeTrim@0#1@%
      \IN@0 @\the\Trimtoks@ @%
        \ifIN@
             \SPLIT@0 @\the\Trimtoks@ @\Trimtoks@\Initialtoks@
             \IN@0\the\Terminaltoks@ @ @%
                 \ifIN@
                 \else \Trimtoks@ {FigNameWithSpace}%
                 \fi
        \fi
      }


   \newtoks\pt@ks
   \def \getpt@ks 0.0#1@{\pt@ks{#1}}
   \dimen0=0pt\expandafter\getpt@ks\the\dimen0@

  \newtoks\Realtoks
  \def\Real#1{%
    \dimen2=#1%
      \SPLIT@0\the\pt@ks @\the\dimen2@
       \Realtoks=\Initialtoks@
            }

   \newdimen\Product
   \def\Mult#1#2{%
     \dimen4=#1\relax
     \dimen6=#2%
     \Real{\dimen4}%
     \Product=\the\Realtoks\dimen6%
        }

 \newdimen\Inverse
 \newdimen\hmxdim@ \hmxdim@=8192pt
 \def\Invert#1{%
  \Inverse=\hmxdim@
  \dimen0=#1%
  \divide\Inverse \dimen0%
  \multiply\Inverse 8}

   \def\Rescale#1#2#3{
              \divide #1 by 100\relax
              \dimen2=#3\divide\dimen2 by 100 \Invert{\dimen2}%
              \Mult{#1}{#2}%
              \Mult\Product\Inverse
              #1=\Product}

  \def\Scale#1{\dimen0=\TheScale %
      \divide #1 by  1280 
      \divide \dimen0 by 5120 %
      \multiply#1 by \dimen0
      \divide#1 by 10   
     }


 \newbox\scrunchbox

 \def\Scrunched#1{{\setbox\scrunchbox\hbox{#1}%
   \wd\scrunchbox=0pt
   \ht\scrunchbox=0pt
   \dp\scrunchbox=0pt
   \box\scrunchbox}}

 \def\Shifted@#1{%
   \vbox {\kern-\YShift@
       \hbox {\kern\XShift@\hbox{#1}\kern-\XShift@}%
           \kern\YShift@}}

   %
 \def\cBoxedEPSF#1{{}\leavevmode 
   \ReadNameAndScale@{#1}\ReadEPSFile@ \ReadBdB@x
   \edef\EPSFSpec@{\the\EPSFDirectorytoks@\the\EPSFNametoks@}
     \TrimFigDims@
     \CalculateFigScale@
     \ScaleFigDims@
     \SetInkShift@
     \Real{\FigScale}\edef\FigSc@leReal{\the\Realtoks}%
   \hbox{$\mathsurround=0pt\relax
         \vcenter{\hbox{%
             \FrameSpider{\hskip-.4pt\vrule}%
             \vbox to \Ht@{\parindent=\z@%
                \FrameSpider{\vskip-.4pt\hrule}\vfil
                \hbox to \Wd@{\hfil}%
                \vfil
                \InkShift@{\EPSFSpecial{\EPSFSpec@}{\FigSc@leReal}}%
             \FrameSpider{\hrule\vskip-.4pt}}%
         \FrameSpider{\vrule\hskip-.4pt}}}%
     $}%
    \CleanRegisters@
    }

 \def\tBoxedEPSF#1{\setbox4\hbox{\cBoxedEPSF{#1}}%
     \setbox4\hbox{\raise -\ht4 \hbox{\box4}}%
     \box4
      }

 \def\bBoxedEPSF#1{\setbox4\hbox{\cBoxedEPSF{#1}}%
     \setbox4\hbox{\raise \dp4 \hbox{\box4}}%
     \box4
      }

  \let\BoxedEPSF\cBoxedEPSF

   %

  \def\EPSFxsize{\afterassignment\ForceW@\ForcedDim@@}
      \def\ForceW@{\ForcedDim@true\ForcedHeight@false}

  \def\EPSFysize{\afterassignment\ForceH@\ForcedDim@@}
      \def\ForceH@{\ForcedDim@true\ForcedHeight@true}

  %
 \def\ReadNameAndScale@#1{\IN@0 scaled@#1@
   \ifIN@\ReadNameAndScale@@0#1@%
   \else \ReadNameAndScale@@0#1 scaled\DefaultMilScale @
   \fi}

 \def\ReadNameAndScale@@0#1scaled#2@{
    \Trim@0#1@%
    \EPSFNametoks@\expandafter{\the\Trimtoks@}%
    \FigScale=#2 pt%
     }

 \def\SetDefaultEPSFScale#1{%
      \global\def\DefaultMilScale{#1}}

 \SetDefaultEPSFScale{1000}

  %

 \def \SetBogusBbox@{%
     \global\BdBoxtoks@{ BoundingBox:0 0 100 100 }%
     \global\def\BdBoxLine@{ BoundingBox:0 0 100 100 }%
     }

 \def\ReadEPSFile@{%
     \openin\EPSFile@
       =\the\EPSFDirectorytoks@\the\EPSFNametoks@
     \relax  
  \ifeof\EPSFile@
     \SetBogusBbox@
     \immediate\write16{}%
     \message{ *** EPS FILE  }%
     \message\expandafter{\the\EPSFNametoks@}%
     \message{ NOT FOUND!  }%
     \immediate\write16{}\relax%
  \else
   \begingroup
   \catcode`\%=12\catcode`\:=12\catcode`\\=12
   \NotIn@true
    \loop
      \ifeof\EPSFile@\NotIn@false
        \SetBogusBbox@
        \immediate\write16{}%
        \message{ *** BoundingBox not found in }%
        \message\expandafter{\the\EPSFNametoks@\space *** }%
        \immediate\write16{}%
      \else\global\read\EPSFile@ to \BdBoxLine@
      \fi
      \global\BdBoxtoks@\expandafter{\BdBoxLine@}%
      \IN@0BoundingBox:@\the\BdBoxtoks@ @%
      \ifIN@\NotIn@false\fi%
    \ifNotIn@\repeat
   \endgroup\relax
  \fi
  \closein\EPSFile@
   }

  \def\ReadBdB@x{
   \expandafter\ReadBdB@x@\the\BdBoxtoks@ @}

  \def\ReadBdB@x@#1BoundingBox:#2@{
    \ForeTrim@0#2@%
    \expandafter\ReadBdB@x@@\the\Trimtoks@ @%
   }

  \newtoks\LLXtoks@  
  \newtoks\LLYtoks@

  \def\ReadBdB@x@@#1 #2 #3 #4@{
      \Wd@=#3bp\advance\Wd@ by -#1bp%
      \Ht@=#4bp\advance\Ht@ by-#2bp%
       \Wd@@=\Wd@ \Ht@@=\Ht@ 
       \LLXtoks@={#1}\LLYtoks@={#2}
      \ifPSOrigin\XShift@=-#1bp\YShift@=-#2bp\fi
     }

   %
  \def\SetEPSFDirectory{
           \bgroup\catcode`\:=12\relax
           \def\G@bbl@##1{}\def\\{}
           \global\edef\B@ckslash{\expandafter\G@bbl@\string\\}
           \global\let\bs\B@ckslash\relax
           \SetEPSFDirectory@}

 \def\SetEPSFDirectory@#1{
    \Trim@0#1@
    \global\EPSFDirectorytoks@\expandafter{\the\Trimtoks@ }\relax
    \egroup}

  %
 \def\TrimTop#1{\advance\TT@ by #1}
 \def\TrimLeft#1{\advance\LT@ by #1}
 \def\TrimBottom#1{\advance\BT@ by #1}
 \def\TrimRight#1{\advance\RT@ by #1}

 \def\TrimFigDims@{%
    \advance\Wd@ by -\LT@
    \advance\Wd@ by -\RT@ \RT@=\z@
    \advance\Ht@ by -\TT@ \TT@=\z@
    \advance\Ht@ by -\BT@
    }

  %
  \def\ForceWidth#1{\ForcedDim@true
       \ForcedDim@@#1\ForcedHeight@false}

  \def\ForceHeight#1{\ForcedDim@true
       \ForcedDim@@=#1\ForcedHeight@true}

  \def\epsfxsize{\afterassignment\ForceW@\ForcedDim@@}
      \def\ForceW@{\ForcedDim@true\ForcedHeight@false}

  \def\epsfysize{\afterassignment\ForceH@\ForcedDim@@}
      \def\ForceH@{\ForcedDim@true\ForcedHeight@true}

  \def\CalculateFigScale@{%
     \ifForcedDim@\FigScale=1000pt
           \ifForcedHeight@
                \Rescale\FigScale\ForcedDim@@\Ht@
           \else
                \Rescale\FigScale\ForcedDim@@\Wd@
           \fi
     \fi}

  \def\ScaleFigDims@{\TheScale=\FigScale
      \ifForcedDim@
           \ifForcedHeight@ \Ht@=\ForcedDim@@  \Scale\Wd@
           \else \Wd@=\ForcedDim@@ \Scale\Ht@
           \fi
      \else \Scale\Wd@\Scale\Ht@
      \fi
      \ForcedDim@false
      \Scale\LT@\Scale\BT@  
      \Scale\XShift@\Scale\YShift@
      }

 \def\HideReservedBoxes{\gdef\FrameSpider##1{}}
 \def\ShowReservedBoxes{\gdef\FrameSpider##1{##1}}

  \ShowReservedBoxes

 \def\hSlide#1{\advance\XSlide@ by #1}
 \def\vSlide#1{\advance\YSlide@ by #1}

  \def\SetInkShift@{%
            \advance\XShift@ by -\LT@
            \advance\XShift@ by \XSlide@
            \advance\YShift@ by -\BT@
            \advance\YShift@ by -\YSlide@
             }
  \def\InkShift@#1{\Shifted@{\Scrunched{#1}}}

   %
  \def\CleanRegisters@{%
      \globaldefs=1\relax
        \XShift@=\z@\YShift@=\z@\XSlide@=\z@\YSlide@=\z@
        \TT@=\z@\LT@=\z@\BT@=\z@\RT@=\z@
      \globaldefs=0\relax}


 \def\SetTexturesEPSFSpecial{\PSOriginfalse
  \gdef\EPSFSpecial##1##2{\relax
    \edef\specialthis{##2}%
    \SPLIT@0.@\specialthis.@\relax
    \special{illustration ##1 scaled
                        \the\Initialtoks@}}}

  \def\SetUnixCoopEPSFSpecial{\PSOrigintrue 
   \gdef\EPSFSpecial##1##2{%
      \dimen4=##2pt
      \divide\dimen4 by 1000\relax
      \Real{\dimen4}
      \edef\Aux@{\the\Realtoks}%
      \includegraphics{##1}}}

  \def\SetRokickiEPSFSpecial{\PSOrigintrue
   \gdef\EPSFSpecial##1##2{%
      \dimen4=##2pt
      \divide\dimen4 by 10\relax
      \Real{\dimen4}
      \edef\Aux@{\the\Realtoks}%
      \includegraphics{##1}}}

  \def\SetOzTeXEPSFSpecial{\PSOriginfalse 
  \gdef\EPSFSpecial##1##2{
     \special{##1\space
       ##2 1000 div \the\mag\space 1000 div mul
       ##2 1000 div \the\mag\space 1000 div mul scale
       \the\LLXtoks@\space neg \the\LLYtoks@\space neg translate
             }}}


 \def\SetArborEPSFSpecial{\PSOriginfalse 
   \gdef\EPSFSpecial##1##2{%
     \edef\specialthis{##2}%
     \SPLIT@0.@\specialthis.@\relax 
     \special{ps: epsfile ##1\space \the\Initialtoks@}}}

 \def\SetClarkEPSFSpecial{\PSOriginfalse 
   \gdef\EPSFSpecial##1##2{%
     \Rescale {\Wd@@}{##2pt}{1000pt}%
     \Rescale {\Ht@@}{##2pt}{1000pt}%
     \special{dvitops: import
           ##1\space\the\Wd@@\space\the\Ht@@}}}


  \def\SetBeebeEPSFSpecial{
   \PSOriginfalse
   \gdef\EPSFSpecial##1##2{\relax
    \special{language "PostScript",
      literal "##2 1000 div ##2 1000 div scale
       \the\LLXtoks@\space neg \the\LLYtoks@\space neg translate",
              overlay "##1"}}}

 \def\SetStandardEPSFSpecial{%
   \gdef\EPSFSpecial##1##2{%
     \immediate\write16{}
     \immediate\write16{%
       **** Sorry! There is still no standard for \string%
       \special \space EPSF integration *****}%
     \immediate\write16{%
      --- So you will have to identify your driver using a command}%
     \immediate\write16{%
      --- of the form \string\Set...EPSFSpecial, in order to get}%
     \immediate\write16{%
      --- your graphics to print.  See BoxedEPSF.doc.}%
     \immediate\write16{}
     \KillEPSFSpecial
     }}

  \def\KillEPSFSpecial{\gdef\EPSFSpecial##1##2{}}

  \SetStandardEPSFSpecial 

 \let\wlog\wlog@ld 

 \catcode`\@=\CatAt
 \catcode`\:=\CatColon

%
%
%
%

\documentstyle{amsppt}
\magnification 1200

\SetRokickiEPSFSpecial 
\SetEPSFDirectory{./}
\SetDefaultEPSFScale{480}
\HideReservedBoxes

\def\figno#1{\botcaption{Figure #1}
\endcaption}

\def\today{31 March 1995} 

\def\End{\operatorname{End}}
\def\Hom{\operatorname{Hom}}
\def\tr{\operatorname{tr}}

\def\H#1 #2 #3 {\Hom_A\left(#2,#1\otimes #3\right)} 
\def\F#1 #2 #3 {\Hom_\field
\left(#2,#1\otimes #3\right)} 

\def\D#1 #2 #3 {\Hom_A\left(#1\otimes #3,#2\right)} 
\def\field{{\Bbb F}}   

\topmatter
\title The equality of\\
 3-manifold invariants \endtitle
\author John W. Barrett and Bruce W. Westbury \endauthor
\date\today\enddate
\address
Department of Mathematics,
University of Nottingham,
University Park,
Nottingham,
NG7 2RD
\endaddress
\email jwb\@maths.nott.ac.uk \endemail
\address
Department of Mathematics,
University of Nottingham,
University Park,
Nottingham,
NG7 2RD
\endaddress
\email bww\@maths.nott.ac.uk \endemail

\abstract
The invariants of 3-manifolds defined by Kuperberg for involutory Hopf
algebras and those defined by the authors for spherical Hopf algebras
are the same for Hopf algebras on which they are both defined.
\endabstract
\endtopmatter

\document
\head {} Introduction \endhead

The purpose of this paper is to compare two previously defined invariants of
3-manifolds.

Let $A$ be a finite-dimensional Hopf algebra over a field $\field$ with
antipode $S$. Then if $S^2=1$ the Hopf algebra is said to be involutory.
Let $A$ be involutory and the dimension of $A$ in the field
$\field$ be not zero. Then it follows from \cite{Larson and Radford 1987}
that $A$ is semisimple and cosemisimple.

For each such Hopf algebra $A$, Kuperberg \cite{1990} has defined an
invariant of closed oriented 3-manifolds. For a manifold $M$, this invariant
is denoted $K(M)$.

The present authors defined an invariant of closed oriented 3-manifolds for
each such Hopf algebra $A$ over an algebraically closed field
\cite{Barrett and Westbury 1993; proposition 6.8}\footnote{The hypothesis
there that the field has characteristic zero can be replaced by the hypothesis
that the algebra has non-zero dimension.}. This invariant is denoted $Z(M)$ for
a manifold $M$ and is called a state sum invariant.

The result of this paper is

\proclaim{Theorem}
Let $A$ be a finite dimensional involutory Hopf algebra over an
algebraically closed field, with $\dim A\ne 0$. Then
$$K(M)=Z(M) \dim A.$$
for all $M$.
\endproclaim

This result implies the following relationship between the scope of the
two invariants. The state sum invariants of Barrett and Westbury \cite
{1993} are defined
for the more general notion of a finite semisimple spherical category of
non-zero dimension. This generalises the notion of the category
of representations of a semisimple Hopf algebra. As we showed in
that paper, examples can be constructed from Hopf algebras which are not
themselves semisimple, such as the quantised universal enveloping algebras.
For these non-semisimple Hopf algebras it is apparently necessary
to use the category theory, as a quotient category has to be taken.
Kuperberg's invariants are examples of the state sum invariants for
which the category theory is not required.

Kaplansky's conjecture is that a finite-dimensional
semisimple Hopf algebra is involutory. This has been proved for
characteristic zero \cite {Larson and Radford 1987}.
If Kaplansky's conjecture is correct, then the state sum invariants which
are defined directly from Hopf algebras without the category theory
are all examples of Kuperberg's invariants if a particular
element of the Hopf algebra, the spherical element, is taken to be 1.
There are further examples with other choices of spherical element.

\head The invariants\endhead

The definitions of the two invariants are given here, followed by the
proof of the theorem stated in the introduction. The definitions rely
on various arbitrary choices; the proof that the invariants do not
depend on these choices is not repeated here, but the reader may find these
in the original references.

Before starting with
the definitions, there are some preliminaries on Hopf algebras.
Let $A$ be a finite dimensional involutory Hopf algebra of non-zero
dimension over the algebraically closed field $\field$. If $a$ is a module
over $A$, the dimension of $a$ is likewise the vector space dimension
of $a$ regarded as an element of the field.
The algebra $A$ can also be regarded as a left $A$-module, and $\dim A$
is unambiguous, the dimension of this module and the algebra $A$ being the
same.
The matrix trace for module $a$ is denoted $\tr_a$. Thus
$$\dim a=\tr_a (1).\tag1$$

The semisimplicity of $A$ has a number of consequences which follow.
Let $I$ be complete set of inequivalent irreducible
left $A$-modules with non-zero dimension. Thus
$$ \dim A = \sum_{a\in I} \left(\dim a\right)^2\ne 0\in \field.\tag 2$$
Also, if $a\in I$ and $b$ is any left $A$-module, then the pairing
$$\align\Hom_A(a,b)&\times \Hom_A(b,a)\to \field\\
(\alpha &,\beta)\mapsto\tr_a\beta\alpha\tag3\endalign$$
is non-degenerate. In this paper, the composition of map $\alpha$ with
$\beta$ is written $\beta\alpha$. This is the opposite convention to that
of Barrett and Westbury \cite {1993}.

The algebra $A$ possesses a unique left integral $i\in A$ such that
the counit $\epsilon$ gives $\epsilon(i)=1$ \cite{Larson and Sweedler 1969}.
For any left $A$-modules $a$ and $b$, $\Hom_\field(a,b)$ is also a left
$A$-module, an element $x\in A$ acting by
$$x\colon\phi\mapsto\sum\Delta_{(1)}(x)\phi S\left(\Delta_{(2)}(x)\right),
\tag 4$$
where the coproduct is written
$$\Delta(x)=\sum\Delta_{(1)}(x)\otimes\Delta_{(2)}(x).\tag5$$
The left integral $i$ has the property that $i^2=\epsilon(i) i$ in any
Hopf algebra. In the semisimple case this gives $i^2=i$, and $i$ acts by
projection in an $A$-module. For a non-semisimple Hopf algebra this would give
$i^2=0$ and the following considerations would not apply.
The subspace $\Hom_A(a,b)\subset\Hom_\field(a,b)$ is exactly the image of
the projection given by the left action of the integral $i$.

The Hopf algebra $A$ has a unique left cointegral $c\in A^*$ with
$c(1)=1$. This is given by
$$c(x)={1\over \dim A} \tr_{A}(x)=
{1\over\dim A}\sum_{a\in I}\dim a\tr_a(x)\tag6$$
\cite {Larson and Radford 1987}, where the first trace is over the left
regular representation of $A$.

\subhead The state sum invariant   \endsubhead

The definition of the state sum invariant given here is the definition
of Barrett and Westbury \cite {1993} specialised to the case of
involutory Hopf algebras with the spherical element equal to 1.

Let $M$ be an oriented closed 3-manifold, and fix a triangulation of $M$.

It is necessary to pick a simplicial structure for the triangulated
manifold $M$. This means that a total ordering is chosen for the vertices
of each simplex such that the orderings of faces are compatible. A suitable
such structure can be obtained by totally ordering all the vertices of $M$.

Let $E$ be the set of edges (1-simplexes) of $M$ and $v$ the number of
vertices of $M$. A labelling of $M$
is a map $l\colon E\to I$. For each labelled simplicial manifold $(M,l)$,
a number $Z(M,l)\in\field$ is defined, and the invariant is determined
from this data by a sum
$$Z(M)=\left(\dim A\right)^{-v} \sum_{l\colon E\to I} Z(M,l)
\prod_{e\in E}\dim(l(e)),\tag 7$$
over the set of all labellings.

It remains to define $Z(M,l)$. For each triangle in $M$, there are three
elements $e_{12},e_{02},e_{01}$ of $I$ assigned to the edges $12,02,01$.
This determines a vector space, $\H e_{12} e_{02} e_{01} $. The vector
space $V(M,l)$ is defined
to be the tensor product of each of these spaces over the set of all
triangles in $M$.

For each tetrahedron in $M$, denote the labelling of the edge
$(ij)$ by $e_{ij}$, for $i<j=0,\ldots 3$. The ordering of the vertices
determines an orientation of the simplex which either agrees (positive) with
that of the manifold or does not (negative). These two cases are considered
separately.

Consider first the case where the simplex is positive.
There are maps
$$\multline \H e_{23} e_{03} e_{02} \otimes \H e_{12} e_{02} e_{01}
\\ \to \H e_{23} e_{03} {e_{12}\otimes e_{01}}
\endmultline\tag 8$$ and
$$\multline  \H e_{23} e_{13} e_{12} \otimes \H e_{13} e_{03} e_{01}
\\ \to \H e_{23} e_{03} {e_{12}\otimes e_{01}} .
\endmultline \tag 9$$
If $A$ is semisimple, then these maps give isomorphisms
$$\multline \bigoplus_{e_{02}} \H e_{23} e_{03} e_{02} \otimes \H e_{12} e_{02}
e_{01}
\\ \to \H e_{23} e_{03} {e_{12}\otimes e_{01}}
\endmultline\tag 10$$ and
$$\multline \bigoplus_{e_{13}} \H e_{23} e_{13} e_{12} \otimes \H e_{13} e_{03}
e_{01}
\\ \to \H e_{23} e_{03} {e_{12}\otimes e_{01}} .
\endmultline \tag 11$$
Comparing these two decompositions gives a map
$$\multline \H e_{23} e_{03} e_{02} \otimes \H e_{12} e_{02} e_{01}
\\ \to \H e_{23} e_{13} e_{12} \otimes \H e_{13} e_{03} e_{01} .
\endmultline \tag 12$$

Our definition, which follows, will give a map which is proportional to this
one, in fact by dividing by $\dim(e_{13})$. This definition is the one used in
Barrett and Westbury \cite {1993}. The constant of proportionality is given
there by lemma 5.4.

The definition uses the fact that the trace gives a non-degenerate bilinear
form, but without necessarily assuming directly that $A$ is semi-simple.

Note that if
$$\align\beta\otimes\delta&\in \H e_{23} e_{03} e_{02} \otimes \H e_{12} e_{02}
e_{01} \\
\alpha\otimes\gamma&\in\Hom_A(e_{23}\otimes e_{12},e_{13})\otimes\Hom_A(
e_{13}\otimes e_{01},e_{03}) \endalign$$
then
$$\align
\gamma(\alpha\otimes 1)&\colon e_{23}\otimes e_{12} \otimes e_{01}\to e_{03}\\
(1\otimes \delta)\beta&\colon e_{03}\to e_{23}\otimes e_{12}\otimes e_{01}.
\endalign$$
Therefore a map \thetag{12} is uniquely determined by the condition that
$\beta\otimes\delta$ maps to the unique element whose pairing with
$\alpha\otimes\gamma$ is
$$\tr\left(\gamma(\alpha\otimes 1)(1\otimes \delta)\beta\right),\tag 13$$
using the non-degenerate bilinear form.

If the simplex is negative, a map
$$\multline
\H e_{23} e_{13} e_{12} \otimes \H e_{13} e_{03} e_{01}
\\ \to
\H e_{23} e_{03} e_{02} \otimes \H e_{12} e_{02} e_{01}
\endmultline$$
is defined by mapping $\alpha\otimes\gamma$ to the unique element which
has trace
$$\tr\left(\beta(1\otimes \delta)(\alpha\otimes 1)\gamma\right)\tag 14$$
with $\beta\in\D e_{23} e_{03} e_{02} $ tensored with $\delta\in
\D e_{12} e_{02} e_{01} $, for all $\beta,\delta$.

The tensor product of the maps \thetag{13} and \thetag{14} over the set of all
tetrahedra in the manifold is a linear map $V(M,l)\to V^\pi(M,l)$,
where $V^\pi(M,l)$ is defined
in the same way as $V(M,l)$ but with the factors permuted by some
permutation $\pi$. There is a unique standard linear map
$V^\pi(M,l)\to V(M,l)$ given by iterating the standard twist
$P\colon x\otimes y\mapsto y\otimes x$. This defines an element of
$\End V(M,l)$, and $Z(M,l)$ is defined to be the trace of this linear map.

\subhead The Kuperberg invariant \endsubhead

Kuperberg's invariant \cite{Kuperberg 1990} is defined by assigning a
number to a Heegaard diagram
for an oriented closed 3-manifold. For the purposes of this paper it is
sufficient to give a definition for the particular Heegaard diagrams
determined by a triangulation.

The triangulation of $M$ determines a Heegaard splitting for $M$ by a
regular neighbourhood of the 1-skeleton. This can be done in a standard way
by taking a subcomplex
of the second barycentric subdivision of $M$. The triangulation determines
standard circles on the Heegaard surface; there is one upper circle for
each edge of the triangulation and one lower circle for each face. The lower
circles are defined to be the intersection of the face with the Heegaard
surface and the upper circles are defined to be the intersection of
the dual face in the dual skeleton with the Heegaard surface. The Heegaard
surface together with the upper and lower circles determines the manifold
up to isomorphism and is called a Heegaard diagram for the manifold.

An arbitrary orientation and a distinguished point is chosen for each circle.
Starting at the distinguished point and travelling in the direction of
the orientation, this gives an ordering of the crossing points between upper
and lower circles which occur around a given circle. The set of upper
circles and the set of lower circles are also regarded as ordered sets.
This endows the set of all crossing points with two distinct total orderings;
one is obtained by traversing all the upper circles in order, the other
by traversing all the lower circles in order.

The Heegaard surface is oriented as the boundary of the thickened 1-skeleton.
This determines a sign for each crossing point as positive if there is
a positively oriented triangle $(012)$ embedded in the Heegaard surface
with vertex $0$ at the
crossing point, $1$ along the upper circle in the direction of its
orientation and $2$ along the lower circle. Otherwise it is negative.

Let $l,u$ be the number of lower and upper circles and $n$ the number of
crossing points. Each lower circle has exactly three crossing points. The
element
$$k=(\Delta\otimes 1)\Delta(i)\in  A\otimes A\otimes A \tag15$$
is associated with each lower circle, with one factor associated to
each crossing point. For each crossing point, define the linear map
$\phi\colon A\to A$ to be the identity if the crossing point is positive
and the antipode $S$ if the crossing point is negative. The linear
functional
$$T\colon(x_1,x_2,\ldots)\mapsto c(\phi(x_1)\phi(x_2)\ldots)$$
on $A\otimes A\otimes \ldots$ is associated to each upper circle, with
one factor associated to each crossing point, using the $\phi$ appropriate
for each crossing point.

A number $\lambda\in\field$ is determined by applying the permutation which
takes the total order determined by the lower circles to the total order
determined by the upper circles to the element
$$\otimes^l k\in \otimes^n A$$
followed by the linear functional $\otimes^n A\to\field$ obtained by
tensoring the $T$ for each upper circle. As before, the permuations act
in the linear space $\otimes^n A$ by iterates of the standard twist map
$P$.

The Kuperberg invariant $K(M)$ is defined to be
$$ K(M)=\lambda \left(\dim A\right)^{u-v+1}.\tag16$$
The exponent $u-v+1$ is the genus of the Heegaard surface.

\subhead The proof of the theorem \endsubhead

Since the edges and upper circles correspond, a labelling
assigns a module in $I$ to each upper circle. In the definition of $\lambda$
in the invariant $K(M)$, replace each $T$ for each upper circle with
the functional
$$(x_1,x_2,\ldots)\mapsto\tr_a(\phi(x_1)\phi(x_2)\ldots),$$
$a$ being the module assigned to the given upper circle.
Denote this element of $\field$ by $\lambda(M,l)$.
Then, using \thetag 6, $K(M)$ is given by the formula
$$K(M)=\left(\dim A\right)^{1-v}\sum_{l\colon E\to I}\lambda(M,l)
\prod_{e\in E}\dim\left(l(e)\right).\tag 17$$
Comparing this with the definition \thetag{7} of the state sum invariant,
to prove the
theorem it remains to show that $Z(M,l)=\lambda(M,l)$ for each labelling $l$.

\goodbreak\topinsert
\centerline{\BoxedEPSF{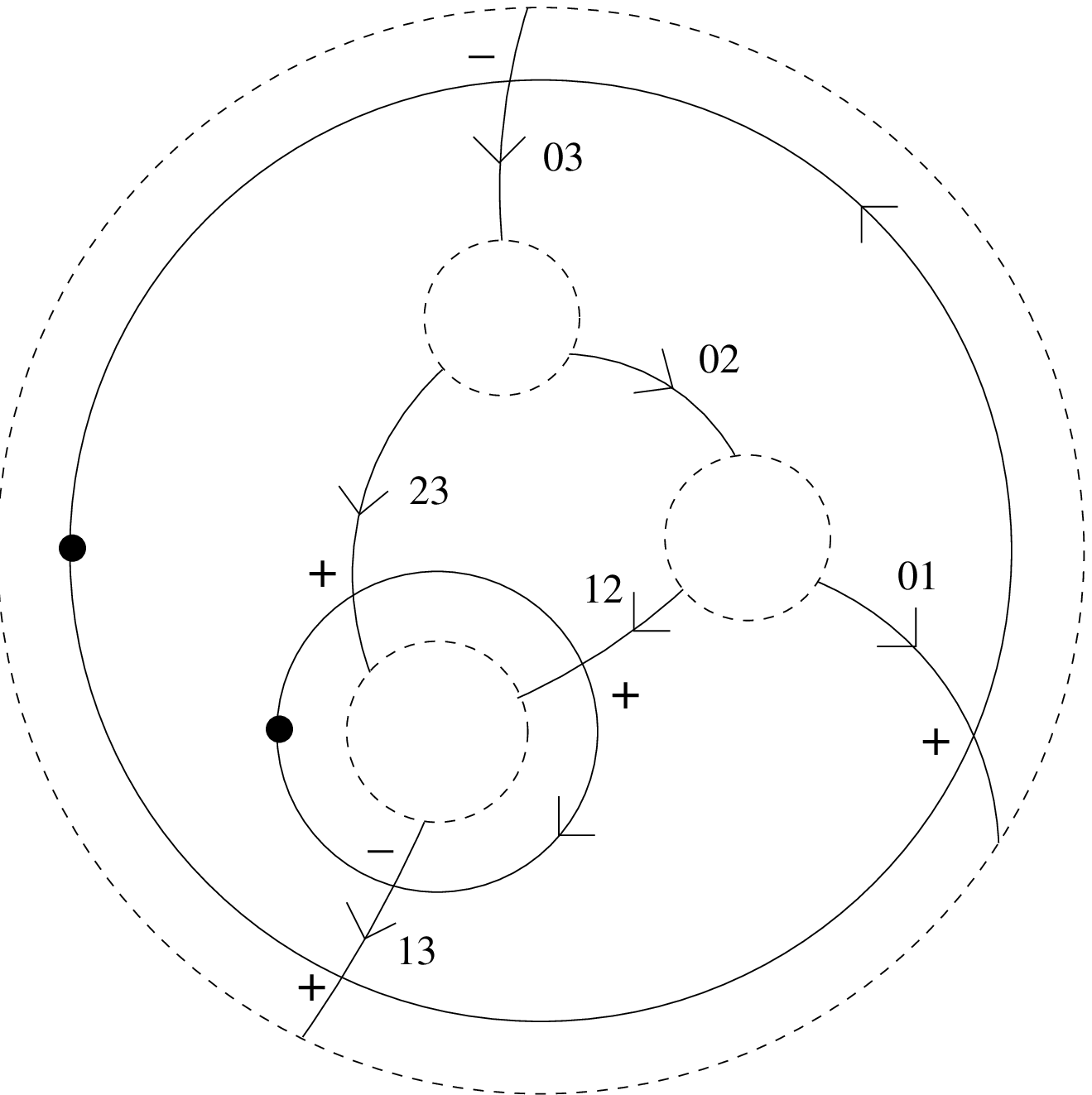}}
\figno 1\endinsert

The Heegaard surface can be split into the union of its intersection with
each tetrahedron in the manifold. The lower circles are on the boundary of each
of these pieces; to simplify the proof they are slid sideways a small amount
so that they lie in a single tetrahedron. The rule for doing this is that the
lower circle lies in the tetrahedron for which the face which corresponds to
it is positively signed in the boundary of the tetrahedron. This is
according to the convention
$$\partial{+(0123)}=+(123)-(023)+(013)-(012),\tag 18$$
and with the opposite signs for $-(0123)$.
Also, the orientations of the circles are chosen with respect to
the orderings of the vertices of $M$. The ordering of the vertices gives
an orientation for each edge of the triangulation, and using the orientation
of the manifold, a rule can be chosen to orient the upper circles in a
consistent way. These circles can be chosen to circulate the edges in
a clockwise or anti-clockwise sense throughout the manifold.
This choice is fixed by the orientations indicated in Figure 1,
in which the intersection of the Heegaard surface with a positively
oriented tetrahedron $+(0123)$ is shown. This is a disk with three punctures.
The dotted line is the boundary,
while the solid line segments are the upper circles, labelled with the
corresponding edge, and the solid line circles are
lower circles.  The signs $+,-$ on the diagram indicate the signs of the
crossing points, which are determined by the orientations. A point is
marked on each lower circle as the distinguished point.
Figure 2 is the same diagram for a negatively oriented tetrahedron.

\goodbreak\midinsert
\centerline{\BoxedEPSF{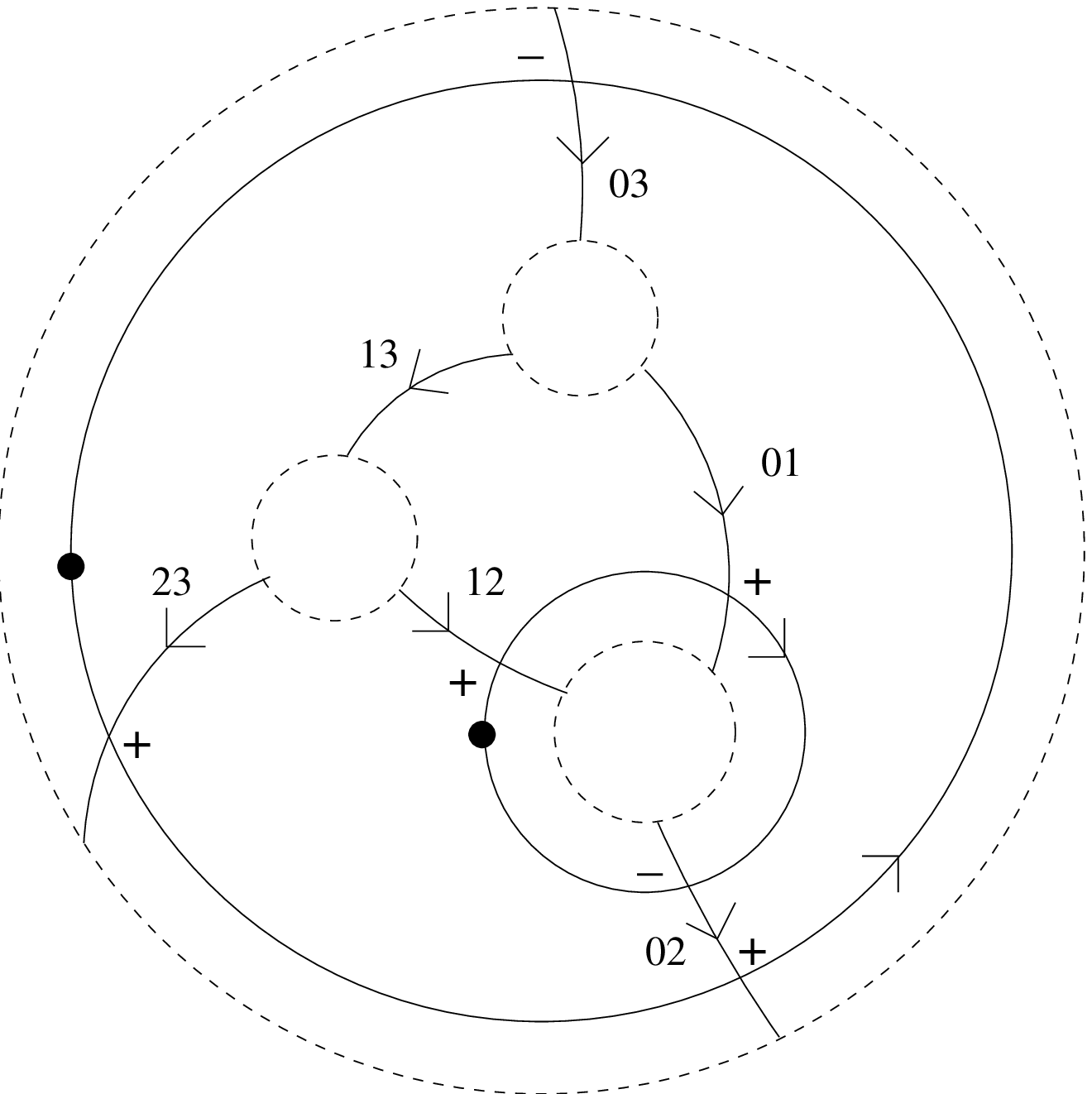}}
\figno 2\endinsert

Each lower circle in Figures 1 and 2 has been chosen so that the signs
of the crossing points are $(+,+,-)$, as in Figure 3. Figure 3 shows part
of a Heegaard surface, with the upper circles labelled by $a$,$b$ and $c$.
The number $\lambda(M,l)$ is calculated by
$$tr_{a\otimes b\otimes c}\bigl(((1\otimes 1\otimes S)(k))\eta\bigr),\tag 19$$
where $k$ is defined in \thetag{15} and
$$\eta\in\End_\field(a\otimes b\otimes c)$$
is the element determined in the definition of $\lambda(M,l)$ for the rest
of the Heegaard diagram, outside Figure 3. Thus $\lambda(M,l)$ can be
considered to be the value of a certain linear functional acting on the
element $\eta$.

\goodbreak\midinsert
\centerline{\BoxedEPSF{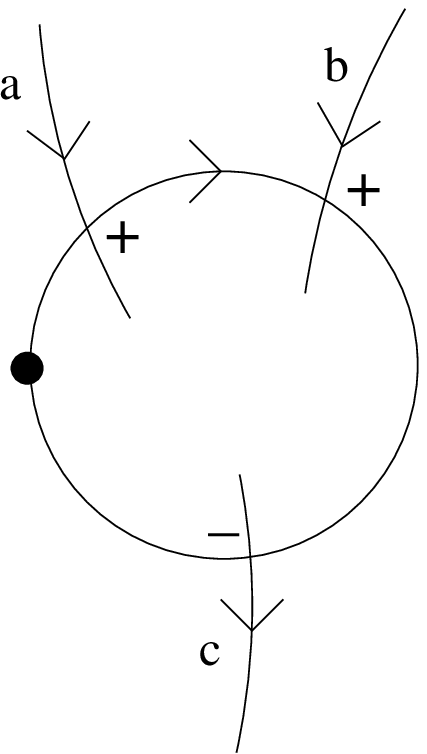}}
\figno 3\endinsert

For fixed elements
$\alpha\in\Hom_\field(a\otimes b,c)$, $\beta\in\Hom_\field(c,a\otimes b)$,
there is a rank one endomorphism of $\Hom_\field(c,a\otimes b)$ defined
by
$$x\mapsto\bigl(\tr_c(\alpha x)\bigr)\beta.\tag 20$$
The matrix trace of this endomorphism is $\tr_c(\alpha\beta)$, and the
set of all such endomorphisms for all $\alpha$ and $\beta$ spans this linear
space of endomorphisms, using the hypothesis that $A$ is semisimple, and
\thetag3.  There is an isomorphism
$$\End_\field\left(\Hom_\field(c,a\otimes b)\right)
\to \End_\field(a\otimes b\otimes c)  $$
determined by
$$\beta\tr_c(\alpha\,\cdot\,)\mapsto P (\alpha\otimes\beta),\tag21$$
with $P$ the usual twist map $c\otimes (a\otimes b)\to (a\otimes b)\otimes c$.
Using this isomorphism, the linear functional
on  $\End_\field\left(\Hom_\field(c,a\otimes b)\right)$ corresponding
to \thetag{19} is given by
$$   \beta\tr_c(\alpha\,\cdot\,)\mapsto\tr_c\left(\alpha\Delta_{(1)}(i)
\beta S(\Delta_{(2)}(i))\right).\tag22$$
This is the trace on the vector space $\Hom_\field(c,a\otimes b)$ with
the map
$$\beta\mapsto  \Delta_{(1)}(i)  \beta S(\Delta_{(2)}(i)). \tag23$$
As explained following \thetag5, this map is a projection on
$\Hom_A(c,a\otimes b)$.

This method can be repeated for the whole of Figure 1, using the
isomorphism of
$$\End_\field(e_{01}\otimes e_{02}\otimes e_{03}\otimes e_{12}\otimes e_{13}
\otimes e_{23})$$
with
$$\multline\Hom_\field\bigl(
 \F e_{23} e_{03} e_{02} \otimes \F e_{12} e_{02} e_{01} ,
\\  \F e_{23} e_{13} e_{12} \otimes \F e_{13} e_{03} e_{01} \bigr).
\endmultline$$
Since each lower circle determines a projection on $\H b a c $, each
space $\F b a c $ can be replaced by $\H b a c $.
A short calculation shows that the part of the Heegaard diagram in Figure 1
determines the same linear map as that associated to the positive
tetrahedron in the state sum invariant. A similar calculation can be
carried out for the negative tetrahedra in the manifold. This shows that
$\lambda(M,l)=Z(M,l)$.

\enddocument